\documentclass[aps,prd,superscriptaddress,nofootinbib,floatfix,preprintnumbers,12pt]{revtex4-2}

\pdfoutput=1

\usepackage{graphicx}
\usepackage{multirow}
\usepackage{enumitem}
\usepackage{amssymb}
\usepackage{amsmath}
\usepackage{xcolor}
\usepackage{xspace}
\usepackage[normalem]{ulem} 

\usepackage{hyperref}

\usepackage{booktabs} 

\newcommand{\nn}{\nonumber}
\newcommand{\beq}{\begin{equation}}
\newcommand{\eeq}{\end{equation}}
\newcommand{\beqa}{\begin{eqnarray}}
\newcommand{\eeqa}{\end{eqnarray}}

\RequirePackage{xspace}
\def\Bbar    {\kern 0.18em\overline{\kern -0.18em B}{}\xspace}

\def\Kbar    {\kern 0.18em\overline{\kern -0.18em K}{}\xspace}
\def\Kb      {\ensuremath{\Kbar}\xspace}

\newcommand{\KorKbar}{\raisebox{7.7pt}{$\scriptscriptstyle(\hspace*{9.9pt})$}
  \hspace*{-13.4
pt}\Kbar{}^{\,\, 0}\,}

\def\OMIT#1{{}}

\makeatletter
\g@addto@macro\bfseries{\boldmath}
\let\Hy@backout\@gobble
\makeatother

\begin{document}
\preprint{CERN-TH-2025-047, CHIBA-EP-272, IPPP/25/46}

\title{CP violation in $K\to\mu^+\mu^-$ with and without time dependence through a tagged analysis}

\author{Giancarlo D'Ambrosio}
\email{gdambros@na.infn.it}
\affiliation{INFN-Sezione di Napoli, Complesso Universitario di Monte S. Angelo, Via Cintia Edificio 6, 80126 Napoli, Italy}

\author{Avital Dery}
\email{avital.dery@cern.ch}
\affiliation{CERN, Theoretical Physics Department, Geneva, Switzerland}

\author{Yuval~Grossman}
\email{yg73@cornell.edu}
\affiliation{Department of Physics, LEPP, Cornell University, Ithaca, NY 14853, USA}

\author{Teppei Kitahara}
\email{kitahara@chiba-u.jp}
\affiliation{Department of Physics, Graduate School of Science, Chiba University, Chiba 263-8522, Japan}
\affiliation{Kobayashi-Maskawa Institute for the Origin of Particles and the Universe, Nagoya University, Nagoya 464-8602, Japan}

\author{Radoslav Marchevski}
\email{radoslav.marchevski@epfl.ch}
\affiliation{École Polytechnique Fédérale de Lausanne, Route de la Sorge, 1015 Écublens, Switzerland}

\author{Diego Martínez Santos}
\email{diego.martinez.santos@cern.ch}
\affiliation{Centro de Investigaci\'on en Tecnoglog\'ias Navales e Industriales, Ferrol Industrial Campus, Dr. Vázquez Cabrera, s/n, 15403, Ferrol, Universidade de A Coruña, Spain}

\author{Stefan Schacht}
\email{stefan.schacht@durham.ac.uk}
\affiliation{Institute for Particle Physics Phenomenology, Department of Physics, Durham University, Durham DH1 3LE, United Kingdom}

\begin{abstract}
We point out that using current knowledge of ${\cal B}(K^0_L\to\mu^+\mu^-)$ and $ {\cal B}(K^0_L\to \gamma\gamma)$, 
one can extract short-distance information from the combined measurement of the time-integrated CP asymmetry, $A_{\rm CP}(K^0\to\mu^+\mu^-)$, and of ${\cal B}(K^0_S\to\mu^+\mu^-)$.
We discuss the interplay between this set of observables, and demonstrate that determining ${\rm sign}[A_{\rm CP}(K^0\to\mu^+\mu^-)]$ would eliminate the discrete ambiguity in the Standard Model prediction for ${\cal B}(K^0_L\to\mu^+\mu^-)$.
We then move on to feasibility studies within an LHCb-like setup, using both time-integrated and time-dependent information, employing $K^0$ and $\Kb^0$ tagging methods.
We find that, within an optimistic scenario, the short-distance amplitude, proportional to the CKM parameter combination $|A^2\lambda^5\bar\eta|$, could be constrained by LHCb at the level of about $35\%$ of its Standard Model value, and the discrete ambiguity in ${\cal B}(K^0_L\to\mu^+\mu^-)_{\rm SM}$ could be resolved at more than $3\sigma$ by the end of the high luminosity LHC. 
\end{abstract}

\maketitle

\newpage
\hrule
\tableofcontents
\vskip .2in
\hrule
\vskip .4in


\section{Introduction}
The CKM paradigm is an extremely successful framework, providing a unified scheme to describe a vast number of experimental observables in a consistent way.
Looking ahead, many new endeavors are underway, aiming to improve the precision with which we measure the CKM parameters, and in parallel searching for further novel tests, scrutinizing the CKM paradigm in uncharted territories.
One such territory is that of kaon physics, in which intriguing developments have occurred in the past few years.
An independent cross-check of the CKM paradigm solely from kaon physics would be a unique test of the Standard Model (SM), providing complementary information to the collection of $B$ physics observables.

Recently, the first observation of the rare decay $K^+\to\pi^+\nu\overline\nu$ was declared by the NA62 collaboration~\cite{NA62:2024pjp}, rejecting the background-only hypothesis with a significance of over $5\sigma$. 
This measurement, together with current theory and experimental knowledge of $|\varepsilon_K|$, provides meaningful constraints on the CKM parameters, beginning to form a CKM determination independent of $B$ physics inputs~\cite{Dery:2025pcx}.
More precise measurements and new kaon observables that can strengthen these CKM tests are the main aim of the kaon program today. This ambitious program would require both substantial advancements in experimental capabilities and precise theory predictions~\cite{Lehner:2015jga, Hoferichter:2023wiy,Chao:2024vvl, Lunghi:2024sjy, HIKE:2023ext, RBC:2020kdj, Buras:2014maa, Cirigliano:2011ny, Gisbert:2017vvj, Grossman:1997sk}.

In this work, our focus is on the rare decay $K^0\to\mu^+\mu^-$, which has profited from substantial theory progress in recent years.
A recent analysis within dispersion theory has allowed a better handle on the long-distance physics contributions, which dominate the $K^0_{L,S}\to\mu^+\mu^-$ decay rates~\cite{Hoferichter:2023wiy} . Additionally, lattice QCD efforts have laid the groundwork for lattice calculations of the long-distance effects~\cite{Chao:2024vvl}. 
In this paper, we adopt a different approach. We bypass the challenge of long-distance physics estimations altogether by focusing on observables that are, to a very good approximation, dictated solely by short-distance contributions, and thus are predicted to excellent precision.
We use the fact that interference effects between $K_L$ and $K_S$ decays to a di-muon pair are sensitive exclusively to short-distance contributions and can be used to cleanly extract CKM parameters~\cite{DAmbrosio:2017klp,Chobanova:2017rkj,Endo:2017ums,Dery:2021mct, Brod:2022khx, Dery:2022yqc}.

We first consider the integrated CP asymmetry, $A_{\rm CP}(K^0\to\mu^+\mu^-)$, as an additional observable and address the following question: \textit{what information can be attained if time-dependent measurements are out of experimental reach?}
We find that the combination of the following time-integrated observables,

\begin{equation}\label{eq:timeIntMeas}
    \Big\{{\cal B}(K^0_L\to\gamma\gamma),\,\, {\cal B}(K^0_L\to\mu^+\mu^-),\,\, {\cal B}(K^0_S\to\mu^+\mu^-),\,\, A_{\rm CP}(K^0\to\mu^+\mu^-) \Big\}\, ,
\end{equation}
fully determines the four non-vanishing theory parameters of the $K^0\to\mu^+\mu^-$ system. This includes a purely short-distance amplitude proportional to the Wolfenstein parameter $|\bar\eta|$, and the unknown relative sign between the short- and long-distance amplitudes \cite{Isidori:2003ts,DAmbrosio:2017klp}. The first two observables in Eq.~\eqref{eq:timeIntMeas} are well-measured, with experimental uncertainties below $1\%$ and $2\%$, respectively.
The remaining two quantities have not been observed yet. The LHCb experiment has set an upper bound on ${\cal B}(K^0_S\to\mu^+\mu^-)$~\cite{LHCb:2020ycd}, with prospects to get close to SM sensitivity in the future.

In the first part of this paper, we address the last remaining observable, the CP asymmetry, which has not yet been explored experimentally.
We derive the SM prediction for $A_{\rm CP}(K^0\to\mu^+\mu^-)$ and its relation to the short-distance amplitude and the discrete ambiguity in the SM prediction of ${\cal B}(K^0_L\to\mu^+\mu^-)$ .
In the second part, we use a fast simulation of the LHCb experiment, considering Upgrade I and Upgrade II detector setups. The developed simulation is used to obtain the sensitivity of the upgraded LHCb detectors to the short-distance amplitude and the sign of $A_{\rm CP}(K^0\to\mu^+\mu^-)$. The analysis uses both time-integrated and time-dependent information to extract the parameters of interest.

\section{Setup}

\subsection{Notations and review}

The time-dependent rate for a beam of initial $K^0$ ($\Kb^0$) particles can be written as~\cite{PDG24}
\begin{eqnarray}\label{eq:timeDep}
\frac{1}{{\cal N}}\frac{d\Gamma(\KorKbar \to\mu^+\mu^-)}{dt} \,  = C_L \,e^{-\Gamma_L t} + C_S \,e^{-\Gamma_S t} \pm  2\,C_\text{Int.}\cos(\Delta M_K t-\varphi_0) e^{-\Gamma t}\, ,
\end{eqnarray}
where $t$ is the time in the kaon rest frame, ${\cal N}$ is a normalization factor, $\Gamma = (\Gamma_L+\Gamma_S)/2$, $\Gamma_S$ ($\Gamma_L$) is the $K_S$ ($K_L$) decay width, and $\Delta M_K$ is the $K_L\text{--}K_S$ mass difference ($\Delta M_K>0$). 

As laid out in detail in Refs.~\cite{Dery:2021mct, Dery:2022yqc}, the four experimental parameters
\begin{equation}
    \left\{C_L, \, C_S, \, C_\text{Int.},\, \varphi_0 \right\} \label{eq:exp-params}
\end{equation}
can be expressed by four theory parameters: three magnitudes and a relative phase,
\begin{equation}
\left\{ |A(K_S)_{\ell=0}|,\,\, |A(K_L)_{\ell=0}|, \,\, |A(K_S)_{\ell=1}|, \,\,\arg\big[A(K_S)_{\ell=0}^* \, A(K_L)_{\ell=0}\big]\right\}\,,
\end{equation}
as follows
\begin{align}\label{eq:match-th-exp}
\begin{aligned}
    C_L\,\,\,\, &= \, |A(K_L)_{\ell=0}|^2\, , \\ 
    C_S\,\,\,\, &= \, |A(K_S)_{\ell=0}|^2 + \beta_\mu^2 |A(K_S)_{\ell=1}|^2 \, , 
    \\ 
    C_\text{Int.}\, &=  
    |A(K_S)_{\ell=0}| |A(K_L)_{\ell=0}| \,, \\
    \varphi_0 \,\,\,\,\, &= \, \arg\big[A(K_S)_{\ell=0}^* \, A(K_L)_{\ell=0}\big]\, .
\end{aligned}
\end{align}
Here, we use
\begin{align}
    \beta_\mu = \sqrt{1-\frac{4m_\mu^2}{m_{K^0}^2}}\,, 
\end{align}
and the quantum number $\ell$ refers to the angular momentum of the dimuon system, where
$\ell=0$ ($\ell=1$) corresponds to its CP-odd (CP-even) configuration since the total angular momentum of $\KorKbar$ is zero.
As shown in Ref.~\cite{Dery:2021mct}, the key to the extraction of the short-distance amplitude is the knowledge of $C_{\rm Int.}$ through the relation
\beq \label{eq:SDparam}
\vert A(K_S)_{{\ell}=0}\vert^2 = \frac{C_{\rm Int.}^2}{C_L}\,.
\eeq

We define the convoluted time-integrated rates
\begin{eqnarray}\label{eq:timeInt}
\widetilde\Gamma(\KorKbar \to\mu^+\mu^-)/{\cal N} &=& 
C_L I_L+ C_S I_S \pm  2\,C_\text{Int.}I_{\rm Int.} \,, 
\end{eqnarray}
where
\beqa\label{eq:IntDef}
I_L(t) &\equiv& \int_{0}^{t}F^{\rm eff}(t^\prime)\,e^{-\Gamma_Lt^\prime}dt^\prime \, , \nonumber \\
I_S(t) &\equiv& \int_{0}^{t}F^{\rm eff}(t^\prime)\,e^{-\Gamma_S t^\prime}dt^\prime \, , \nonumber \\
I_{\rm Int.}(t) &\equiv& \int_{0}^{t}F^{\rm eff}(t^\prime) \cos(\Delta M_K  t^\prime-\varphi_0)e^{-\Gamma t^\prime}dt^\prime \, .
\eeqa
In the above, we introduce a positive efficiency function, $F^{\rm eff}(t)$, which encodes the acceptance taking into account all lab-frame variables and cuts folded into a function of proper time.
This function can be determined experimentally using $K_S^0\rightarrow\pi^+\pi^-$ decays. We note that while $I_L$ and $I_S$ do not depend on the experimental parameters of interest in Eq.~(\ref{eq:exp-params}), $I_{\rm Int.}$ depends on $\varphi_0$, which is known experimentally up to a four-fold discrete ambiguity~\cite{Dery:2022yqc}. In the following, we treat $I_{\rm Int.}$ as known and specify the taken value of $\varphi_0$ used for the numerical simulation when applicable.

\subsection{Assumptions and approximations}
\label{sec:assum}
Throughout the paper we use the following assumptions: 
\begin{enumerate}
    \item We neglect CPV in mixing. The ${\cal O}(\varepsilon_K)$ correction to the CP asymmetry in $K^0\to\mu^+\mu^-$ is negligible for current purposes \cite{Brod:2022khx}.  
    
    \item We assume that short-distance physics only generates the $\ell=0$ final state. This is equivalent to assuming no scalar operators are at play, and is fulfilled within the SM to ${\cal O}(m_K^2/m_W^2)\sim 10^{-5}$~\cite{Hermann:2013kca}.

    \item We neglect CPV in the long-distance contributions to $K^0\to\mu^+\mu^-$. This assumption is also fulfilled in the SM to ${\cal O}(10^{-3})$~\cite{Dery:2021mct}.
\end{enumerate}

\section{The time-integrated CP asymmetry}

The time-integrated CP asymmetry in the decay of a neutral kaon into a CP eigenstate $f$ is given by
\begin{equation}
\widetilde{A}_{\rm CP}^f=\frac{ \widetilde D_f }{\widetilde S_f}\, ,
\end{equation}
where $\widetilde D_f(t),\,\widetilde S_f(t)$ are the difference and sum of the convoluted time-integrated decay rates (see Eq.~\eqref{eq:timeInt}),
\begin{align}
    \widetilde D_f &= \widetilde\Gamma(\Kb^0\to f) - \widetilde\Gamma(K^0\to f)\, , \\ \nonumber
    \widetilde S_f &=  \widetilde\Gamma(K^0\to\ f) + \widetilde\Gamma(\Kb^0\to f).
\end{align}
For $f=\mu^+\mu^-$, we can write these in terms of the coefficients of Eq.~\eqref{eq:match-th-exp} as
\begin{align}
	\frac{\widetilde D_{\mu^+\mu^-}(t)}{{\cal N}} &=  
    -4 \,C_{\rm Int.}I_\text{Int.}(t)\,, \\
	\frac{\widetilde S_{\mu^+\mu^-}(t)}{{\cal N}} &= 
    2\big(C_L I_L(t) + C_S I_S(t)\big)\,.
\end{align}
We then have
\begin{equation}\label{eq:ACP}
   \widetilde{A}_{\rm CP}(t) \equiv \widetilde{A}_{\rm CP}^{\mu^+\mu^-}(t) \, = \, -\frac{2\,C_\text{Int.} I_\text{Int.}(t)}{C_L I_L(t) +C_SI_S(t)}\, .
\end{equation}

We denote any quantity that involves convolutions with the efficiency function $F^{\mathrm{eff}}(t)$ 
with a tilde. In the limit $F^{\mathrm{eff}}(t)\equiv 1$,  we have
\begin{align}
\widetilde{A}_{\rm CP} = A_{\rm CP}\,,
\end{align}
where $\widetilde{A}_{\rm CP}$ is the \lq\lq{}experimental\rq\rq{}, and $A_{\rm CP}$ the \lq\lq{}theoretical\rq\rq{} CP asymmetry.

Equation~\eqref{eq:ACP} demonstrates the sensitivity of the integrated CP asymmetry, $\widetilde{A}_{\rm CP}(t)$, to the short-distance amplitude via $C_{\rm Int.}$ and Eq.~\eqref{eq:SDparam}.
The extraction of $C_{\rm Int.}$ relies on knowledge of the remaining parameters in Eq.~\eqref{eq:ACP}: (i) parameters appearing in the integrals of  Eq.~\eqref{eq:ACP}, of which $\varphi_0$ is the only one that is not unambiguously determined; (ii) parameters $C_L$ and $C_S$. Two of the above are straightforwardly related to well-measured branching ratios~\cite{Dery:2022yqc},
\begin{align}\label{eq:CLcosvarphi}
    C_L &= \frac{16\pi m_K}{\beta_\mu \tau_L}\times {\cal B}(K^0_L\to\mu^+\mu^-)\,, \\ \nonumber
    \cos^2\varphi_0 &= \frac{\alpha_{em}^2 m_\mu^2}{2\beta_\mu m_K^2}\log^2\left(\frac{1-\beta_\mu}{1+\beta_\mu}\right)\times \frac{{\cal B}(K^0_L\to\gamma\gamma)}{{\cal B}(K^0_L\to\mu^+\mu^-)}\, ,
\end{align}
containing a four-fold discrete ambiguity in the determination of $\varphi_0$.
Putting this discrete ambiguity aside for the moment leaves us with two parameters in Eq.~\eqref{eq:ACP}, which must be determined experimentally:
\begin{equation}
    \big\{C_{\rm Int.},\,\, C_S\big\}\,.
\end{equation}
These parameters are related to the total $K_S$ branching ratio and the short-distance dominated $\ell=0$ mode as follows~\cite{Dery:2021mct},
\begin{align}
    C_S &= \frac{16 \pi m_K}{ \beta_{\mu} \tau_S } \times \mathcal{B}(K^0_S\rightarrow \mu^+\mu^-)\, , \\ \nonumber
    C_{\rm Int.} &= C_L \sqrt{\frac{\tau_L}{\tau_S} \frac{\mathcal{B}(K^0_S\rightarrow \mu^+\mu^-)_{l=0}}{\mathcal{B}(K^0_L\rightarrow \mu^+\mu^-) }}\,.
\end{align}

Note that there is a way to slightly improve the sensitivity to the time-integrated CP asymmetry. 
When one chooses the value of $\varphi_0$, then the sign of $\cos(\Delta M_K t - \varphi^0)$ is known as a function of the time. 
Using $T_{\varphi_0}$, 
\begin{equation}
   T_{\varphi_0} 
    \simeq \,\begin{cases}
        3.8 \,\tau_S
         \qquad \sin 2 \varphi_0 >0\\ 
        2.9 \, \tau_S
         \qquad \sin 2 \varphi_0 <0\,,
    \end{cases} 
\end{equation}
which is determined by the condition $\cos(\Delta M_K T_{\varphi_0} - \varphi_0)=0$,
one can construct the following improved efficiency function for $I_{\rm Int.}(t)$:
\begin{align}
& I_{\rm Int.}^{\rm improved}(t) = \nn\\ &\begin{cases}
I_{\rm Int.}(t)\,, \quad t \leq T_{\varphi_0} \\
 \int_{0}^{T_{\varphi_0}}F^{\rm eff}(t^\prime) \cos(\Delta M_K  t^\prime-\varphi_0)e^{-\Gamma t^\prime}dt^\prime
- \int_{T_{\varphi_0}}^{t}F^{\rm eff}(t^\prime) \cos(\Delta M_K  t^\prime-\varphi_0)e^{-\Gamma t^\prime}dt^\prime\,,  & t > T_{\varphi_0}\,.
\end{cases}
\end{align}
We find that this artificial efficiency function can amplify ${A}_{\rm CP}(t)$ up to $\sim 30\%$ for the case of $\sin 2 \varphi_0 <0$, while its effect is $\sim 10\%$  for $\sin 2 \varphi_0 >0$. (Note that when one considers $F^{\rm eff}(t)$ for the Upgrade II in Table \ref{tab:acceptances}, $\widetilde{A}_{\rm CP}(t)$ is amplified only by up to $\sim 10\%$  for  $\sin 2 \varphi_0 <0$.)

\subsection{SM prediction for $A_{\rm CP}$}
\label{sec:SMprediction}
\begin{table}[]
    \centering
    \begin{tabular}{c|c}
        \hline\hline
        ~~${\cal B}(K^0_L\to\mu^+\mu^-)_{\rm exp.}$~~ & ~~$(6.84 \pm 0.11)\cdot 10^{-9}$~\cite{PDG24}~~ \\
        ${\cal B}(K^0_L\to\gamma\gamma)_{\rm exp.}$ & $(5.47\pm 0.04)\cdot 10^{-4}$~\cite{PDG24} \\
        \hline
        ~~${\cal B}(K^0_S\to\mu^+\mu^-)_{\rm SM}$~~ & ~~$(5.18 \pm 1.5)\cdot 10^{-12}$~\cite{DAmbrosio:2017klp}~~ \\
        ${\cal B}(K^0_S\to\mu^+\mu^-)_{\ell=0,\rm SM}$ & $(1.7 \pm 0.2)\cdot 10^{-13}$~\cite{Brod:2022khx} \\
        \hline\hline
    \end{tabular}
    \caption{Inputs for SM predictions. \label{tab:Inputs}}  
\end{table}
\begin{figure}[t]
    \includegraphics[width=0.8\textwidth]{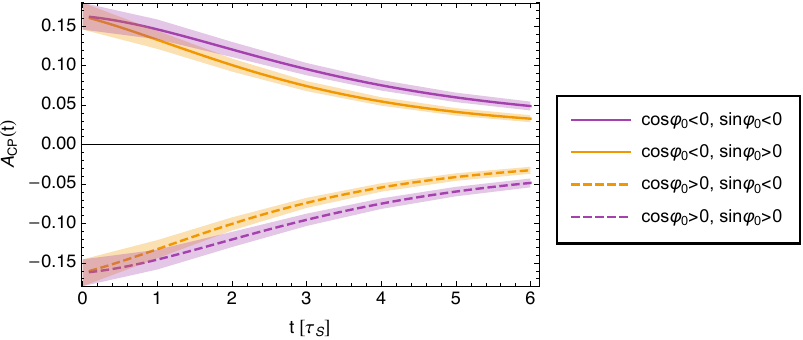}
    \caption{SM prediction for $A_{\rm CP}$ as a function of the end point of the integrals, as in Eq.~\eqref{eq:ACP}. The shaded regions correspond to the $1\sigma$ uncertainties on the predictions. \label{fig:ACPTheory}
    }
\end{figure}

We use the experimental inputs and SM predictions detailed in Table~\ref{tab:Inputs}, to predict the SM CP asymmetry in the limit of $t \to 0$\,:
\begin{equation}
    A_{\rm CP}^{\rm SM}(t\to 0) \, = \,\begin{cases}
        +\, 0.163
        \, \pm \, 0.017\,, \qquad \cos\varphi_0<0\\ 
        -\, 0.163\, \pm \, 0.017\,, \qquad \cos\varphi_0>0.
    \end{cases} 
\end{equation}
Here the total uncertainty is dominated by the theoretical knowledge of ${\cal B}(K_S\to\mu^+\mu^-)$. The knowledge of ${\cal B}(K_S\to\mu^+\mu^-)$, captured by the parameter $C_S$, is limited by the prediction of long-distance contributions.
Figure~\ref{fig:ACPTheory} shows the SM prediction for the integrated CP asymmetry, as a function of the end point of the integrals, $t$. At $t=0$ there is only a two-fold discrete ambiguity. For any other upper integration limit there is a four-fold discrete ambiguity.
However, as the current theory uncertainty on $\widetilde A_{\rm CP}$ is of ${\cal O}(10\%)$, each pair of solutions related by $\sin\varphi_0\to -\sin\varphi_0$ overlap within $2\sigma$. 
We see that the only currently significant discrete ambiguity corresponds to $\rm{sign}(\widetilde A_{\rm CP})$.
The measurement of the sign of the CP asymmetry would resolve the discrete ambiguity in the prediction of $\mathcal{B}(K_L\rightarrow \mu^+\mu^-)$, as we explain in Sec.~\ref{sec:sign-absorptive}.
We also obtain 
\begin{align}
C_{\mathrm{Int.}}/C_L &= 0.119 \pm 0.007\,,\\
C_S/C_L &= 0.4 \pm 0.1\,.
\end{align}

\subsection{Extraction of short-distance parameters}
In the following we demonstrate how, within the assumptions detailed in subsection~\ref{sec:assum}, the short-distance parameter of interest can be extracted from a measurement of $\widetilde A_{\rm CP}$ and ${\cal B}(K_S\to\mu^+\mu^-)$.
From Eq.~\eqref{eq:ACP}, we can isolate the short-distance parameter of Eq.~\eqref{eq:SDparam} as,
\begin{equation}\label{eq:extractSD}
    \frac{C_{\rm Int.}^2}{C_L} \, = \, \widetilde A_{\rm CP}^2\frac{\big(C_L I_L +C_S I_S\big)^2}{ 4\,C_L \, I_{\rm Int.}^2}\, .
\end{equation}
Equation~\eqref{eq:extractSD} can be rewritten in terms of time-integrated rates and the CP asymmetry, %
\begin{equation}\label{eq:extractSD2}
    {\cal B}(K^0_S)_{\ell=0} \, =  \, \frac{1}{4}\,\widetilde A_{\rm CP}^2\cdot {\cal B}(K^0_L) \frac{\tau_S}{\tau_L}  \frac{I_L^2}{I_{\rm Int.}^2} \left(  1 + \frac{\tau_L}{\tau_S}\frac{{\cal B}(K^0_S)}{{\cal B}(K^0_L)}\, \frac{I_S}{I_L}\right)^2 \, , 
\end{equation}
where we use the shorthand notations, ${\cal B}(K^0_S) = {\cal B}(K^0_S\to\mu^+\mu^-)$, ${\cal B}(K^0_L) = {\cal B}(K^0_L\to\mu^+\mu^-)$.
The right-hand side of Eq.~\eqref{eq:extractSD2} involves only known quantities and the two future observables, $\widetilde A_{\rm CP}$ and ${\cal B}(K_S)$.
The left-hand side of Eq.~\eqref{eq:extractSD2} is a theoretical observable, containing the short-distance information. 
Within the SM, it is predicted to ${\cal O}(\varepsilon_K)$ as~\cite{Brod:2022khx} 
\begin{align}\label{eq:SMSD}
    {\cal B}(K^0_S)_{\ell=0} \, =  \, \frac{\beta_\mu \tau_S}{16\pi m_K}\left|\frac{2\,G_F^2 m_W^2}{\pi^2}f_K m_K m_\mu Y_t\right|^2\cdot (\lambda^5 A^2\,\bar\eta)^2 \, ,
\end{align}
indicating that the combined measurement of $\widetilde A_{\rm CP}$ and ${\cal B}(K^0_S)$ would determine the CKM combination $(\lambda^5 A^2 \bar \eta)^2$. The value of the loop function $Y_t$ can be found in Ref.~\cite{Brod:2022khx}. 

From Eqs.~(\ref{eq:extractSD2},~\ref{eq:SMSD}), assuming prior knowledge on the CKM parameters $A,\, \lambda$, we derive the relative uncertainty on the extraction of the parameter $\bar\eta$, arising from the uncertainty in the measurement of $\widetilde A_\text{CP}$ and ${\cal B}(K^0_S)$.
\begin{equation}\label{eq:etabar_uncertainty}
   \frac{\delta \bar\eta}{\bar\eta} \, = \, \sqrt{\left(\frac{\delta \widetilde A_\text{CP}}{\widetilde A_\text{CP}}\right)^2 +  R_{K^0_S}^2\left(\frac{\delta {\cal B}(K^0_S)}{{\cal B}(K^0_S)}\right)^2 }\, , \qquad \text{with} \quad R_{K^0_S} = \frac{\frac{\tau_L}{\tau_S}\frac{{\cal B}(K^0_S)}{{\cal B}(K^0_L)}\frac{I_S}{I_L}}{\left(1+\frac{\tau_L}{\tau_S}\frac{{\cal B}(K^0_S)}{{\cal B}(K^0_L)}\frac{I_S}{I_L}\right)}\, .
\end{equation}
The value of the ratio $R_{K^0_S}$ depends on the efficiency function in the integrals of Eq.~\eqref{eq:IntDef}, as well as on the integration interval. We find that for a range of reasonable values, it obeys $R_{K^0_S}^2<0.1$, implying that, assuming comparable relative errors on $\widetilde A_{\rm CP}$ and ${\cal B}(K^0_S)$, the uncertainty on $\widetilde A_{\rm CP}$ dominates the uncertainty on the extracted value of $\bar\eta$.

\subsection{Determining $\mathrm{sign}[\cos\varphi_0]$ and ${\rm sign}[A(K_L \to \gamma \gamma)]$ \label{sec:sign-absorptive}}
Although the decay rate of $K^0_L\to\mu^+\mu^-$ has been well measured, its SM prediction suffers from the sign ambiguity of the $K^0_L \to \gamma \gamma$ three-point amplitude:
 Its sign determines whether destructive or constructive interference between
the short-distance and the local long-distance contributions to the $K^0_L\to\mu^+\mu^-$ rate, and hence 
two distinct SM predictions for ${\cal B}(K^0_L\to\mu^+\mu^-)$ exist \cite{Hoferichter:2023wiy,DAmbrosio:2017klp, Isidori:2003ts}.
This discrete ambiguity is directly related to the discrete ambiguity in the determination of the phase-shift $\varphi_0$, appearing in the integral $I_{\rm Int.}$ (defined in Eq.~\eqref{eq:IntDef}).

Since $C_{\rm Int.},\, C_L,\, C_S,\, I_L$ and $I_S$ are all positive by definition, the sign of the integrated CP asymmetry, $\widetilde A_{\rm CP}$ as in Eq.~\eqref{eq:ACP}, is determined by the sign of $I_{\rm Int.}$.
Using $\cos^2\varphi_0 = 0.96$ and the value of $\Delta M_K$, we find 
\begin{equation}
	|\cos\Delta M_K t \cos\varphi_0| \, > \, |\sin\Delta M_K t \sin\varphi_0| \qquad \text{for all}\quad t\lesssim 2\tau_S\,,
\end{equation}
which implies that the sign of $I_{\rm Int.}$ is in turn determined by the sign of $\cos\varphi_0$ throughout the whole integration range of interest, $t\in[0,2\tau_S]$.
We deduce that once the sign of $\widetilde A_{\rm CP}$ is measured, 
the discrete ambiguity in the determination of $\cos\varphi_0$ will be resolved.
The sign of $A(K^0_L \to \gamma \gamma)$ is then determined by the sign of the short-distance amplitude \cite{Dery:2022yqc}, as follows 
\begin{equation}
    {\rm sign}[\widetilde A_{\rm CP}] \,= \, -{\rm sign}[\cos\varphi_0] \, = \,  {\rm sign}[\sin\theta_{\rm SD}]\, {\rm sign}[C_{\rm had.}]\,,
\end{equation}
where 
the detailed definitions of $\theta_{\rm SD}$ and $C_{\rm had.}$\! are given in Ref.~\cite{Dery:2022yqc}. 
Note that  ${\rm sign}[C_{\rm had.}]$ is related to the unknown sign of the $K^0_L \rightarrow \gamma \gamma$ amplitude: 
\begin{equation}
\operatorname{sign}\left[A\left(K^0_L \rightarrow \gamma \gamma\right)\right]= -{\rm sign}[G_8-G_{27}]\,  {\rm sign}[C_{\rm had.}] \,\operatorname{sign}\left[A\left(K^0_L \rightarrow \pi^0 \rightarrow \gamma \gamma\right)\right]\,,
\end{equation}
where $G_{8(27)}$ denotes the leading coupling of the $|\Delta S| =1$ chiral nonleptonic weak Lagrangian.
Corresponding lattice determinations can be found in Refs.~\cite{Blum:2015ywa,RBC:2020kdj}.

In the SM, 
since $\sin\theta_{\rm SD} > 0$ (Eq.~(4.23) of \cite{Dery:2022yqc}), we have 
\begin{align}
 {\rm sign}[\widetilde A_{\rm CP}] = {\rm sign}[C_{\rm had.}] \,,
\end{align}
so one can determine the unknown overall sign of $A(K^0_L \to \gamma \gamma)$ from a measurement of the CP asymmetry \cite{DAmbrosio:2017klp}.
This sign has important implications for the SM prediction of $\mathcal{B}(K^0_L\rightarrow \mu^+\mu^-)$ that has two-fold evaluations depending on $\text{sign}[C_{\text{had.}}]$ \cite{Hoferichter:2023wiy}\footnote{We acknowledge Bai-Long Hoid for supplying these numbers in private communication.} 
\begin{align}\label{eq:prediction-KLmumu}
\mathcal{B}\left(K^0_L \rightarrow \mu^{+} \mu^{-}\right)_{\rm{SM}}=\left\{\begin{array}{l}
7.44^{+0.41}_{-0.34}  \times 10^{-9}\qquad \text{for}~~ \widetilde A_{\rm CP}>0\,, \\
6.83^{+0.24}_{-0.17}  \times 10^{-9}\qquad \text{for} ~~\widetilde A_{\rm CP}<0\,,
\end{array}\right.
\end{align}
and for the resulting bounds on physics beyond the SM, see \emph{e.g.}~Ref.~\cite{Chobanova:2017rkj}. Note that the assignment in Eq.~(\ref{eq:prediction-KLmumu}) depends on the overall phase of the long-distance contributions to $\mathcal{B}\left(K^0_L \rightarrow \mu^{+} \mu^{-}\right)$.

Let us point out 
the possibility that measuring the sign of  $A(K^0_L \to \gamma \gamma)$ may shed light on other problems where also the same ${\cal O} (p^4)$ cancellation of $\pi^0$ and $\eta$ pole contributions appears: $A(K^0_L \to \pi^+ \pi ^-  \gamma)$ \cite{Ko:1990jz,Hoferichter:2023wiy} and the $K^0_L - K^0_S$ mass difference \cite{Donoghue:1986ti}; unfortunately also in these problems the required ${\cal O} (p^6)$ calculation is not free of uncertainties and so the possibility to have a firm result is welcome.

\section{LHCb as a case study --- prospects and requirements}
\label{sec:LHCb}

We investigate the sensitivity of a tagged analysis to the parameters of interest defined in Eq.~(\ref{eq:exp-params}) with the LHCb upgraded detector. The flavor of the $\KorKbar$ at production can be estimated by searching for a companion charged kaon in $pp \rightarrow K^0K^-X$ processes \cite{DAmbrosio:2017klp}. 
To estimate the sensitivity to the parameters of interest, we need to determine:
\begin{itemize}
    \item $K^0$ and $\Kb^0$ yields produced at the LHC. The $K^0$ and $\Kb^0$ are produced in similar amounts, corresponding to a cross section times multiplicity of 0.3 barn for each of them. There is a small, momentum-dependent, production asymmetry of about a few-percent. We neglect it in our study, assuming that it can be estimated from data through $K^0\rightarrow\pi\mu\nu$ decays.
    
    \item $K^0\rightarrow\mu^+\mu^-$ detector efficiency and decay-time acceptance. We obtain the efficiency and decay-time acceptance by using the fast simulation software described in Ref.~\cite{Chobanova:2020vmx}.
    
    \item Tagging power for $K^0$. The tagging power, hereafter $T_P$, is defined as $T_P = \varepsilon_T D^2$, where $\varepsilon_T$ is the efficiency of the tagging algorithm to provide a response -- either right or wrong -- and $D$ is the dilution. 
    
    \item The expected signal significance, $S/\sqrt{S + B}$. It is important to notice that here $S$ includes not only $K_S^0\rightarrow\mu^+\mu^-$ but the full $K^0(\Kb^0)\rightarrow\mu^+\mu^-$ yield, which is in total more than twice the $K_S^0$ yield alone.
\end{itemize}

\subsection{Detector acceptance, scenarios}

To first approximation, the decay-time acceptance for $K^0\rightarrow\mu^+\mu^-$  can be described by an allowed decay-time range (driven by selection cuts) and an exponential decay factor~\cite{AlvesJunior:2018ldo} 

\begin{equation}\label{eq:FeffApprox1}
    F^{\rm eff}(t) \, \approx \, \Theta(t-t_{\min})\Theta(t_{\max}-t)\,\,e^{-\beta_{acc.}t}\, ,
\end{equation}
\noindent where $\beta_{acc}$ was $\approx86$ $ns^{-1}$ in Run-I but its value is very dependent on the kinematic requirements of the trigger and selection. Thus, we recalculate it with fast simulation for the conditions of a highly efficient trigger. In addition, the Phase-II upgrade is expected to replace the current Upstream Tracker (UT) by a pixel detector, the Upstream Pixel (UP). The UP will provide a much better resolution of the dimuon opening angle (and hence of the invariant mass) for kaon decays downstream the VeloPix (VP), and thus $K^0\rightarrow\mu^+\mu^-$ decays downstream of the VP could be well separated from $K_S^0\rightarrow\pi^+\pi^-$ decays. This is very difficult to do with the current (and past) detector, and is the reason why only decays inside the VELO/VP volume are used. 
In our fast simulation, we find that downstream $K^0$ decays reconstructed using the UP have a mass resolution nearly as good as those decaying inside the VP, thus potentially having comparable $S/B$. In terms of acceptance effects, the use of downstream $K^0\rightarrow\mu^+\mu^-$ decays would roughly double the sample size, as well as further reducing $\beta_{acc}$ and increasing $t_{\max}$. Thus, we define two acceptance scenarios: one in which only kaons decaying inside the VP volume are used (hereafter LL, since both muons are reconstructed as long tracks), and one in which also downstream kaons are used (LL+DD, since both muons can be reconstructed either as long tracks or as downstream tracks). The data from LHCb Upgrades Ia and Ib will most likely be only LL, while the LHCb Upgrade~II data~\cite{LHCb:2021glh} could optimistically aim at the LL+DD scenario.  The values of $t_{\max}$, $\beta_{acc}$ and the yield increase are shown in Table \ref{tab:acceptances}. 
The proper time resolution is completely negligible in comparison to $\Delta M_K$ and is neglected throughout this paper.

\begin{table}[t!]
    \centering
    \vspace{1em}
    \begin{tabular}{c|c|c|c|c}\hline\hline
        & ~Scenario~ & ~$S/S_{LL}$~ & $t_{\max}$ & $\beta_{acc}$\\
        \hline
      Upgrade~I &  LL       & 1 & ~$\sim220$ $ps$ ~& ~$21.1$ $ns^{-1}$ ~\\
        \hline   
      Upgrade~II &  LL+DD       & $\approx 2-3$ & $\sim500$ $ps$ & $4.1$ $ns^{-1}$ \\\hline\hline
    \end{tabular}
    \caption{Definition of scenarios for the detector acceptance. The LL scenario refers to the case in which only long tracks can be used, and corresponds to our default assumptions for Upgrade~I. The LL+DD scenario refers to the case in which downstream $K^0\rightarrow \mu^+\mu^-$ decays are reconstructed with comparable precision to those made of long tracks and corresponds to our default assumptions for Upgrade~II. \label{tab:acceptances}}
\end{table}
\subsection{Flavor Tagging}

The tagging power depends on two quantities: the tagging efficiency (the probability to tag a $K^0$ even if the assigned flavor is wrong), and the mistag rate, $\omega$, which is the probability of the assigned tag to be wrong:
\begin{equation}\label{eq:TP}
T_P = \varepsilon_T D^2 = \varepsilon_T(1-2\omega)^2\,.
\end{equation}
We consider that the $K^0$ production flavor can be obtained by searching for a companion charged kaon, in a similar way as the Same Side Kaon (SSK) tagger works for $B_s^0$ mesons~\cite{LHCb:2012zja}. To estimate the tagging power, we include in the fast simulation the charged kaons and pions of the underlying event. The pions are included because they can be misidentified as kaons and thus contribute to the mistag rate. The fast simulation software in~\cite{Chobanova:2020vmx} does not include the Particle Identification System. Thus, in our study we assume a $3\%$ probability for a pion to be misidentified as a kaon.

We first test our setup simulating the SSK in $B_s^0\rightarrow J/\psi(\rightarrow\mu^+\mu^-)\phi(\rightarrow K^+K^-)$ simulated decays. We find a tagging power of $\approx3\%$, which is in the ballpark of SSK's performance in LHCb's full simulation and data.

\begin{figure}[t]
    \includegraphics[width=0.6\textwidth]{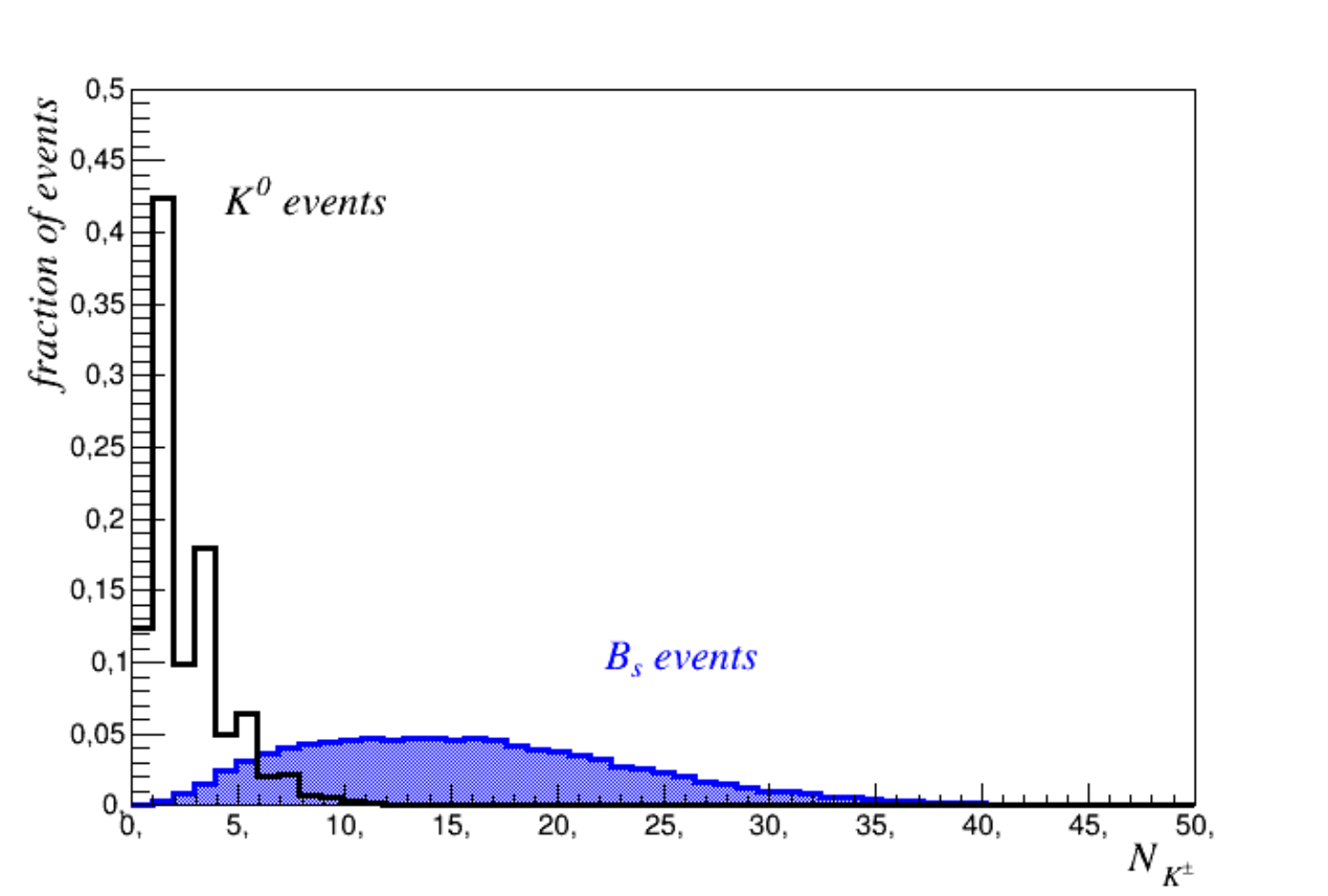}\hfill
    \caption{Kaon multiplicity per $pp$ simulated collision for $pp\rightarrow K^0 X$ events (black lines) and $pp\rightarrow B_s^0 X$ events. The simulations are done with \texttt{PYTHIA 8.2.2.4}~\cite{Sjostrand:2014zea} using \texttt{SoftQCD:all} settings. It can be seen that $B_s^0$ events have a much higher kaon multiplicity, providing a higher mistag rate than in average $K^0$ events. This explains why we find much better tagging power for an SSK-like algorithm in $pp\rightarrow K^0X$ events compared to $pp\rightarrow B_s^0X$ events.}
    \label{fig:kaon_multiplicity}
\end{figure}

In the simulation of $K^0\rightarrow\mu^+\mu^-$ decays, we found an encouraging dilution $D \approx 60\%$ for a tagging efficiency of $\varepsilon_T\approx62\%$. This results in a tagging power as high as $T_P \approx 22\%$, which is much better than that obtained for $B_s^0$ decays. This is due to the much lower $K^\pm$ multiplicity in $pp\rightarrow K^0(\Kb^0)K^{-(+)}X$ decays compared to  $pp\rightarrow B_s^0(\bar{B_s^0})K^{+(-)}X$, according to our Pythia setup. This is illustrated in Fig. \ref{fig:kaon_multiplicity}. 

We also investigate the use of a companion $\pi^\pm$ or $\Lambda^0$, but find they give subdominant contributions to $T_P$. Further improvements on the tagging power could be achieved
by the use of full event information via deep-learning, as developed for $B$ decays~\cite{Prouve:2024xgq}, but
such a study goes beyond the scope of this paper.

\subsection{Background yield assumptions \label{sec:background}}
For our sensitivity studies, we use signal-only pseudo-experiments. In the presence of background, the sensitivity dilutes significantly, so we estimate an {\it effective} background-free signal yield 
\begin{equation}\label{eq:Seff}
    S_{eff} = \frac{S^2}{S+B}\,.
\end{equation}

\begin{figure}[t]
    \includegraphics[width=0.49\textwidth]{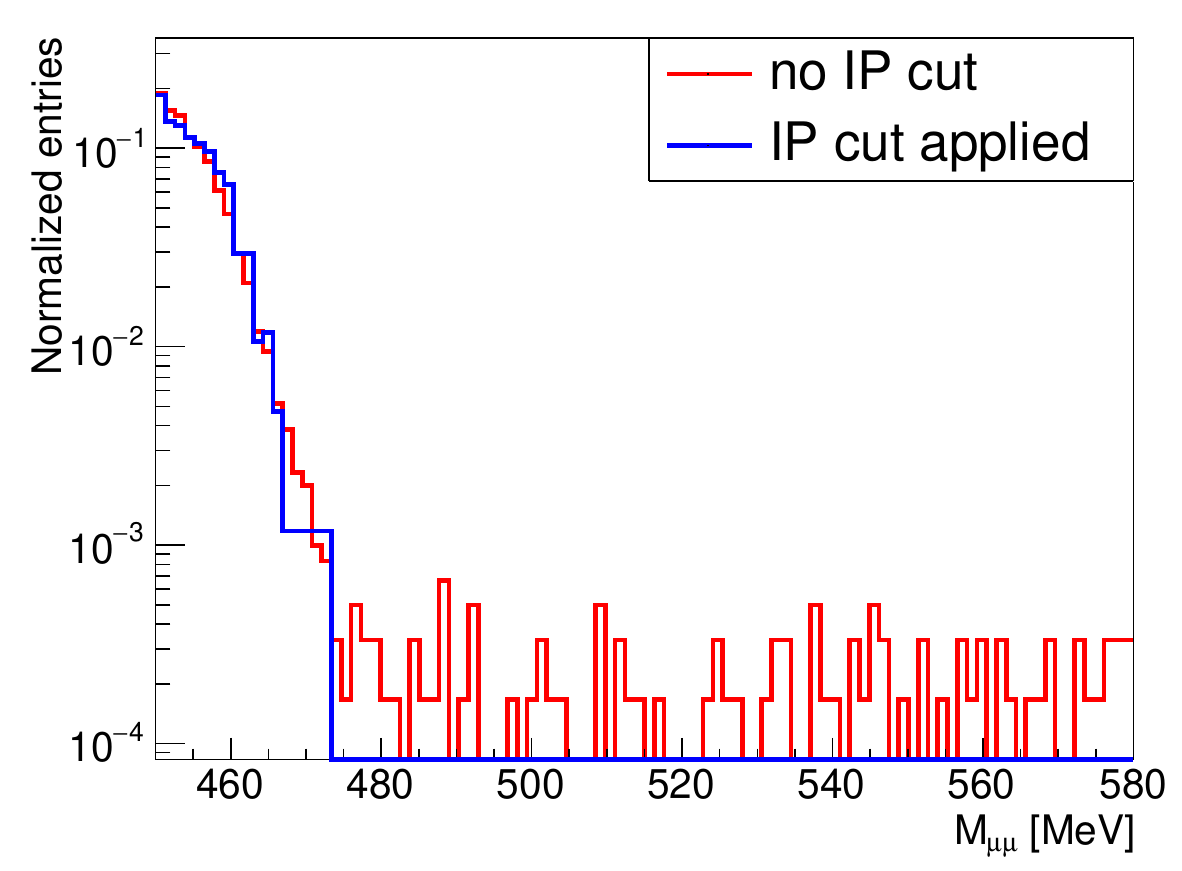}\hfill        \includegraphics[width=0.49\textwidth]{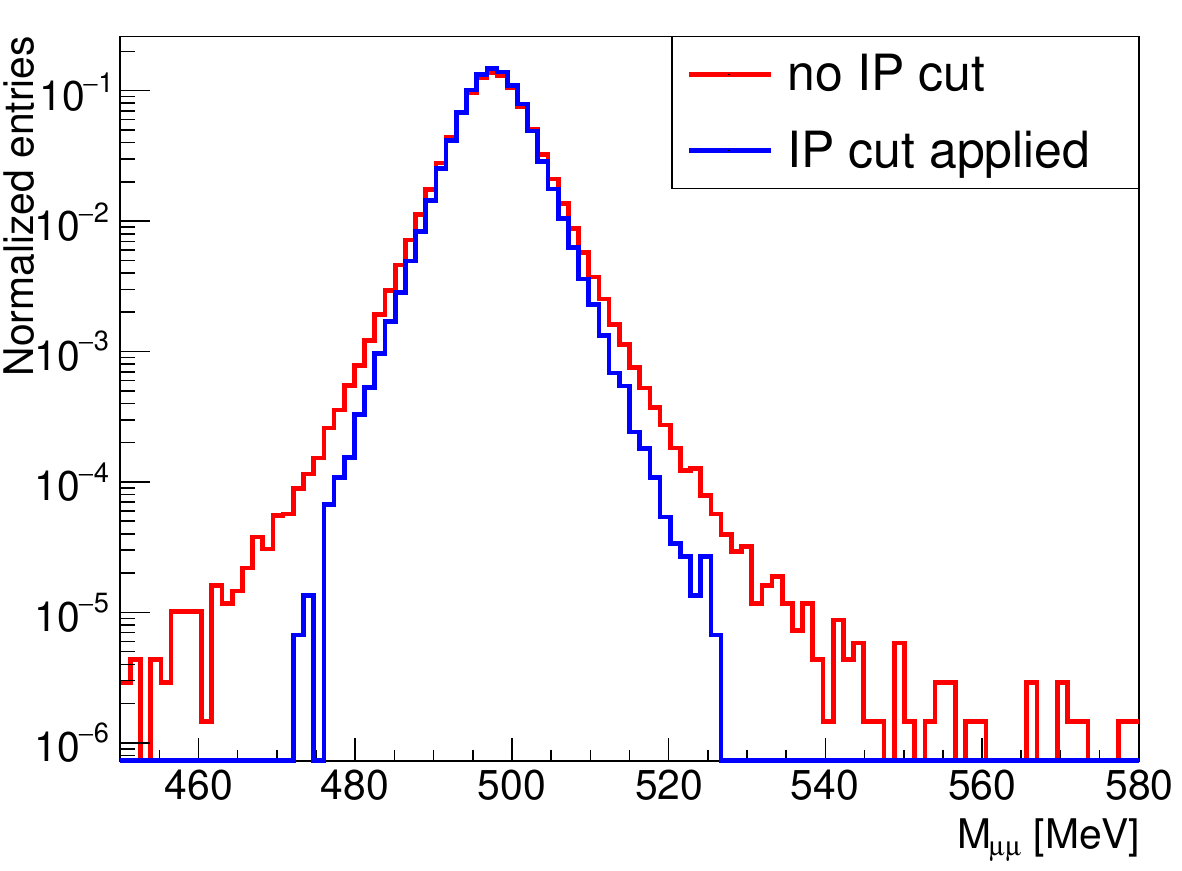}\hfill

    \caption{Invariant mass distributions from fast simulation before (red) and after (blue) applying an impact parameter cut that eliminates half of the signal. Left: $K_S^0\rightarrow\pi^+(\rightarrow\mu^+\nu)\pi^-(\rightarrow\mu^-\bar{\nu})$ decays. Right:  $K_S^0\rightarrow\mu^+\mu^-$ decays. It can be seen that the decays in flight that enter the $K_S^0$ peak get removed almost entirely and, in addition, the mass resolution of the remaining $K_S^0\rightarrow\mu^+\mu^-$ is better than in the unfiltered sample, further improving the signal-background separation.  }
    \label{fig:kaon_mass}
\end{figure}

The current experimental prospects for LHCb Upgrade~II are to reach an upper limit for the $K_S^0\rightarrow \mu^+\mu^-$ branching fraction close to the SM level at $95\%$ CL with $\approx 300~\text{fb}^{-1}$ of integrated luminosity. Based on the invariant mass plots of Ref.~\cite{RamosPernas:2728965} (Appendix E) we approximate it to a single-bin experiment with $\approx 50000$ background events for an expected SM signal yield of 450 $K_S^0\rightarrow \mu^+\mu^-$ decays. The total $K^0(\Kb^0)\rightarrow\mu^+\mu^-$ yield would be about 1900 events once one takes into account the $K_L^0\rightarrow\mu^+\mu^-$ contribution. Thus, $S_{eff} \approx 1900^2/51900 = 69$ events.
However, it has to be noted that further background suppression is possible. The $K_S^0\rightarrow\mu^+\mu^-$ fits of Run-II data suggest that the background is dominated by $K_S^0\rightarrow\pi^+\pi^-$ where at least one of the pions decayed in flight to a muon and a neutrino.  In this study, we simulate decays in flight and find that they can be separated from the signal with cuts in the reconstructed impact parameter of the $K^0$ with respect to the $pp$ collision point. For genuine signal, this impact parameter differs from zero only because of resolution effects. But decays in flight smear the momentum strength of the reconstructed background tracks, and hence the impact parameter of $K_S^0\rightarrow\pi^+(\rightarrow\mu^+\nu)\pi^-(\rightarrow\mu^-\bar{\nu})$ can take wider ranges. This is illustrated in Fig \ref{fig:kaon_mass}, where it can be seen that the $K^0_S\to\pi^+\pi^-$  (with pions decaying in flight) that leak into the $K^0\to\mu^+\mu^-$ signal region can be removed with a cut on the impact parameter of the $K^0$.  For our most optimistic scenarios we consider that the background can be reduced by one or
two orders of magnitude by sacrificing about half of the efficiency.

\subsection{Effective Yield}

The tagging power as well as the signal purity can be condensed into a single effective yield:
\begin{equation}\label{eq:Yeff}
    Y_{eff} = T_P S_{eff} = \varepsilon_T(1-2\omega)^2\frac{S^2}{S+B}\,.
\end{equation}
\noindent With this definition, the sensitivity scales as $1/\sqrt{Y_{eff}}$. Then, we can consider different paths
to achieve a given $Y_{eff}$ and thus a given sensitivity to the interference terms. It has to be noted that
this quantity does not encapsulate the different changes in the decay time acceptance shape (namely the values of $\beta_{acc}$ and $t_{\text{max}}$) due to the presence of the
UP (it only accounts for the yield being doubled). But as we will see later,  the impact of $\beta_{acc}$ and $t_{\text{max}}$ is small. 
Thus, there are different ways to achieve a given $Y_{eff}$, some examples are given in Table~\ref{tab:yeffs}.

\begin{table}[t!]
    \centering
    \vspace{1em}
    \begin{tabular}{c|c|c|c|c|c}\hline\hline
       & $T_P (\%)$ & $S^2/(S+B)$ improvement factor& $S_{eff}^{DD}/S_{eff}^{LL}$ & L [$\text{fb}^{-1}$] & $Y_{eff}$\\
        \hline
     & 22& $\times9$ & 1 & 350 & 319 \\
        Upgrade II  &  22 & $\times9$ & 1.9 & 250 & 330 \\
       &  22& $\times9$ & 1.9 & 300 & 396\\
      &  15 & $\times 6.7$ & 0.7  & 300 & 118 \\
      & 3  & $\times 9$  & 0.7 & 330 & 35 \\
     & 10 & $\times2.2$ & 0 & 300 & 15 \\
   Upgrade I  &  22 & $\times9$ & 0 & 50 & 23 \\\hline\hline
    \end{tabular}
    \caption{Examples of various performance scenarios and their impact on the effective yield, see text for details. The column $S_{eff}^{DD}/S_{eff}^{LL}$ indicates the gain from the downstream sample. This factor would currently be zero, but could be up to $\sim 1 -2$ with the Upstream Pixel, thanks to the improvement on invariant mass resolution. \label{tab:yeffs}}
\end{table}

The baseline expectation for $S_{eff}$ for a given luminosity $L$ is
\begin{align}
S_{eff} &= 69 \times \frac{L}{300\, \text{fb}^{-1}}\,, \label{eq:Seff-expectations}
\end{align}
see Sec.~\ref{sec:background}. In Table~\ref{tab:yeffs} we consider several scenarios where Eq.~(\ref{eq:Seff-expectations}) is multiplied with an improvement factor resulting from additional background suppression as described in Sec.~\ref{sec:background}, as well as a factor 
reflecting the additional gain on top of that from a downstream sample that would be in place in case of the UP being installed as part of Upgrade~II.
The actual $S_{eff}$ in these cases is improved to be 
\begin{align} \label{eq:SeffImproved}
S_{eff} &= 69 \times \frac{L}{300\, \text{fb}^{-1}} \times 
        \left(\text{improvement factor}\right)\times 
        \left(1 + \frac{S_{eff}^{DD}}{S_{eff}^{LL}}\right)\,.
\end{align}
$Y_{eff}$ is then obtained using Eq.~(\ref{eq:Yeff}).

\subsection{Expected sensitivity}
\label{sec:results}

\begin{figure}[t]
    \includegraphics[width=0.6\textwidth]{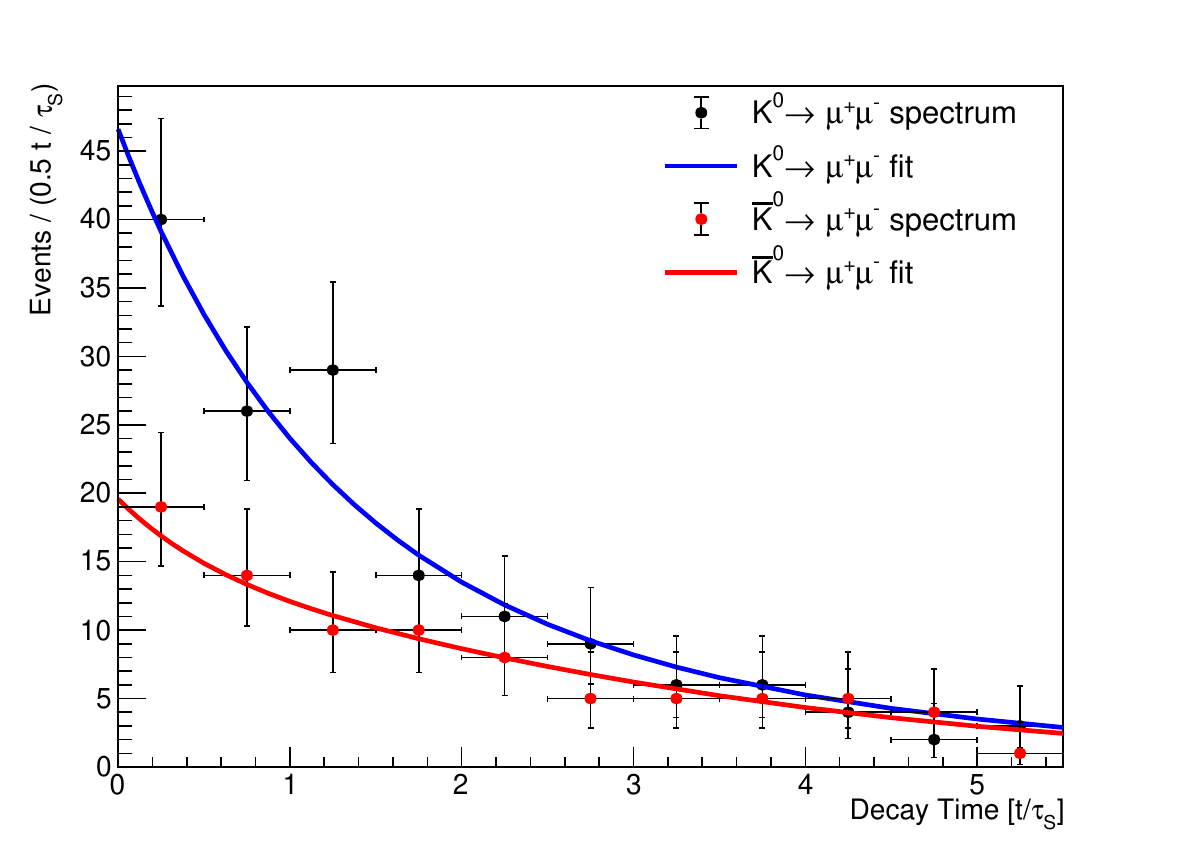}\hfill
    \caption{Decay-time distributions for $K^0\rightarrow\mu^+\mu^-$ (blue) and $\Kb^0\rightarrow\mu^+\mu^-$ decays (red), using the decay-time acceptance for the Upgrade II scenario. The functions after the unbinned simultaneous fit are overlaid. The spectra are shown for $Y_{eff} \sim 300$ events.
    \label{fig:exp_decay-time}}
\end{figure}

In this study, the parameters $C_L$ and $\varphi_0$ are fixed to their experimentally constrained central values. The value of $C_L$ is determined using Eq.~\eqref{eq:CLcosvarphi} and the inputs listed in Table~\ref{tab:Inputs}. For $\varphi_0$, a representative value is chosen in the quadrant where $(\cos{\varphi_0}>0, \sin{\varphi_0}>0)$, leading to negative expected values of $\tilde{A}_{\rm CP}$ (see Fig~\ref{fig:ACPTheory}). We consider two scenarios, corresponding to Upgrade I and Upgrade II, each defined by a different integration range $(t_{\min}, t_{\max})$ and a distinct efficiency function $F^{\text{eff}}(t)$.  These integration ranges are used to evaluate the integrals defined in Eq.~\eqref{eq:IntDef}.

The expected experimental time-integrated CP asymmetry, $\tilde{A}_{\rm CP}$, is computed according to Eq.~(\ref{eq:ACP}) using the SM values for $C_{\text{Int}}$ and $C_S$. Next, we generate the decay-time distribution of $K^0$ and $\Kb^0$ decays into $\mu^+\mu^-$ accounting for the decay-time acceptance $F^{\rm eff}(t)$, for a given $Y_{eff}$. The proportions of $K^0\rightarrow\mu^+\mu^-$ and $\Kb^0\rightarrow\mu^+\mu^-$ decays are determined by the calculated $\tilde{A}_{\rm CP}$ and generated according to a Poisson distribution. The decay time distributions for $Y_{eff}\sim300$  events are shown in Fig.~\ref{fig:exp_decay-time}.

The sensitivity to the CP-violating parameter $|\bar{\eta}|$ is extracted from a simultaneous fit to the $K^0$ and $\Kb^0$ decay-time distributions. An unbinned extended maximum likelihood fit is used, where the yields of $K^0\rightarrow\mu^+\mu^-$ and $\Kb^0\rightarrow\mu^+\mu^-$ explicitly depend on $\tilde{A}_{\rm CP}$, being the main constraint for $\bar{\eta}$. We assume that the value of $C_S$ could be obtained from the LHCb untagged analysis with an ultimate precision of $\mathcal{O}(16\%)$ and we use that as a constraint in our fit, scaling the precision with $Y_{eff}$ as follows
\begin{align}
\frac{\sigma( C_S)}{C_S}  &= \sqrt{\frac{1}{N_{K_S^0}^{eff}}} \leq \sqrt{ \frac{1}{Y_{\mathrm{eff}}}               \frac{\mathcal{B}(K_S^0\to\mu^+\mu^-)\varepsilon_{K^0_S}+\mathcal{B}(K_L^0\to\mu^+\mu^-)\varepsilon_{K^0_L}}{\mathcal{B}(K_S^0\to\mu^+\mu^-)\varepsilon_{K^0_S}}  
                }\,,
\end{align}
\noindent where $\varepsilon_{K_{S(L)}^0}$ represents the efficiency for $K_{S(L)}^0$. This treatment is conservative in the sense that it assumes that the untagged sample size is as big as the tagged sample, while in reality it will be significantly bigger ($S_{eff}-Y_{eff}>Y_{eff}$). Nevertheless, we tried different parameterizations for the precision of the untagged analysis but the impact on our studies is very small.

\begin{figure}[t]
    \includegraphics[width=0.49\textwidth]{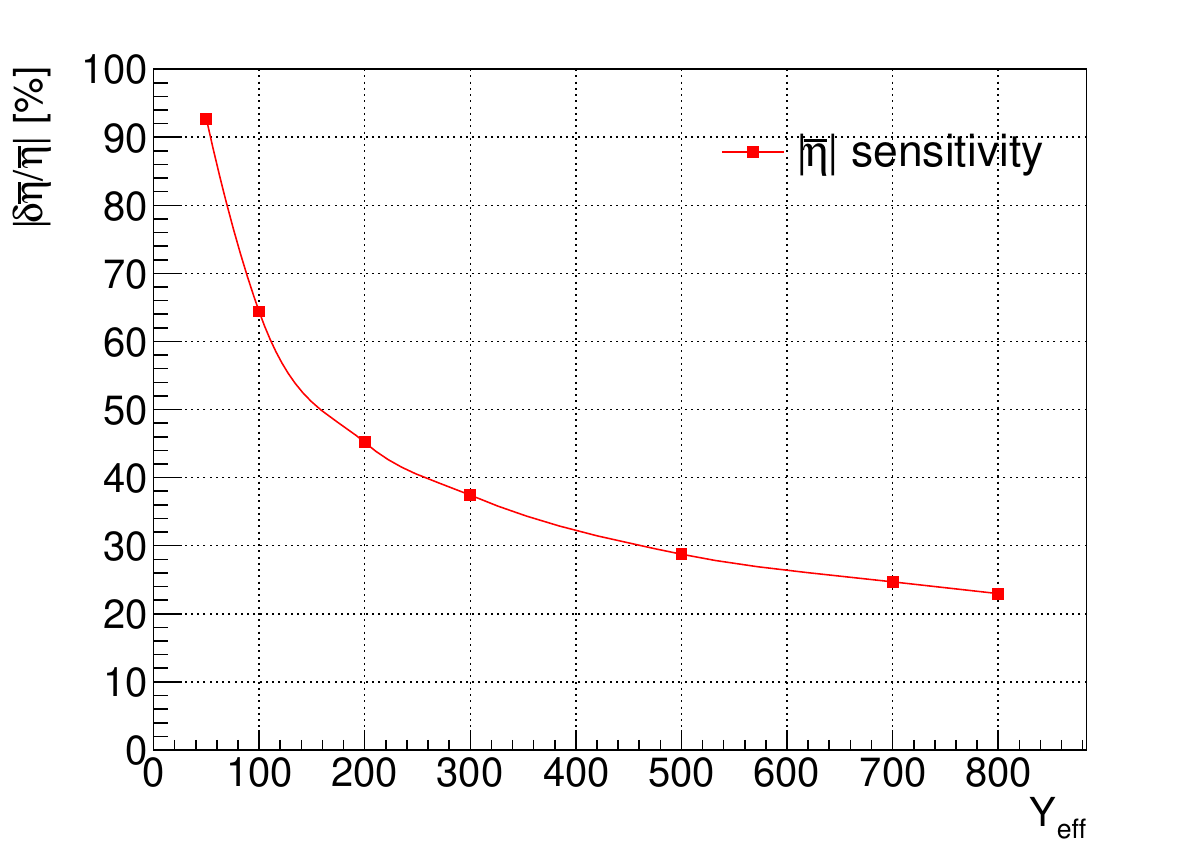}\hfill
    \includegraphics[width=0.49\textwidth]{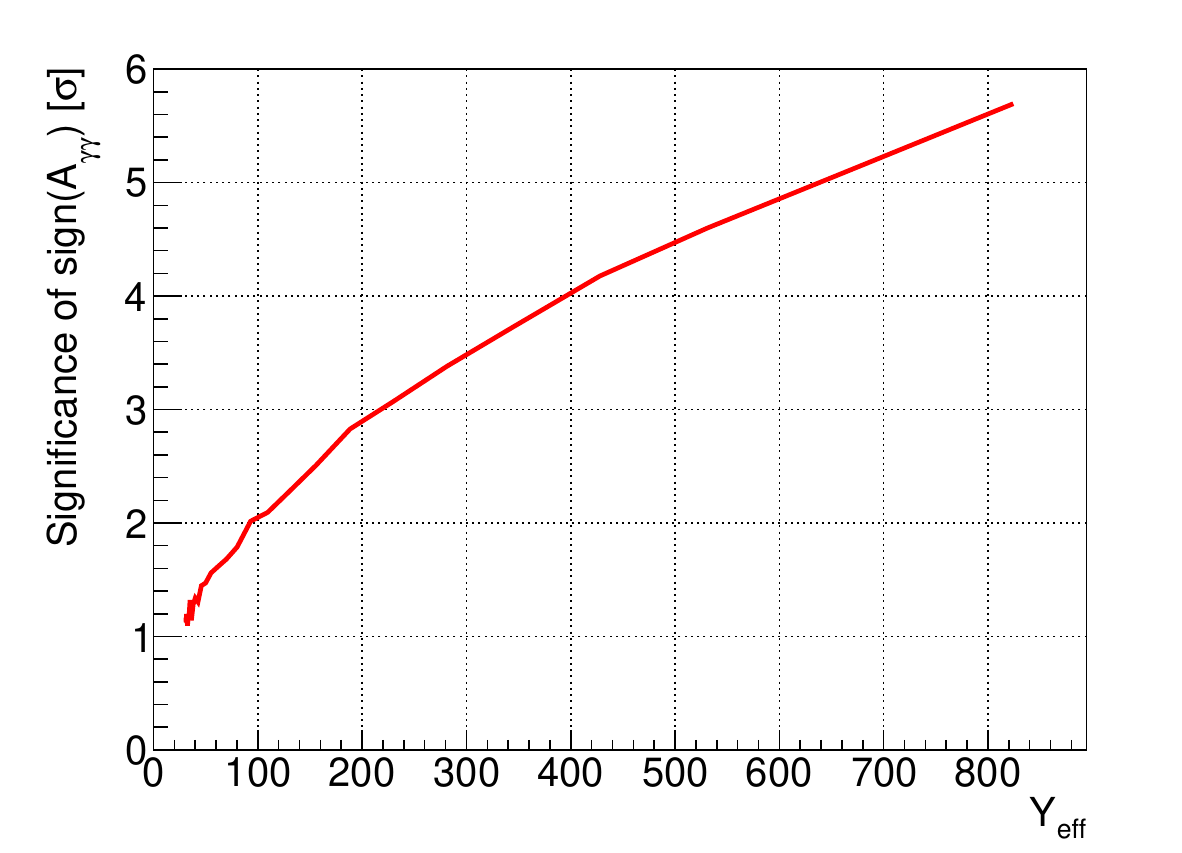}\hfill
    \caption{Experimental sensitivity to  $\vert\bar{\eta}\vert$ (left), and the significance of the sign of $A_{\gamma\gamma}$ assuming the SM (right), as a function of the effective yield, $Y_{eff}$. It can be seen that with $Y_{eff} \sim 300$ events, LHCb can measure $|\bar\eta|$ with uncertainty below $40\%$, and resolve ${\rm sign}[A_{\gamma\gamma}]$ at more than three standard deviations. See Tab.~\ref{tab:yeffs} for various scenarios and the corresponding $Y_{eff}$ values. 
    \label{fig:exp_results}}
\end{figure}

The entire procedure is repeated 10,000 times to demonstrate statistical stability and validate coverage for the extracted parameters. The resulting relative uncertainties on $|\bar{\eta}|$ found by this analysis are presented in Fig.~\ref{fig:exp_results} left, as a function of $Y_{eff}$.

Figure~\ref{fig:exp_results} left demonstrates that LHCb could find evidence for non-zero $\bar{\eta}$ via $\widetilde A_{\rm CP}(K^0\rightarrow\mu^+\mu^-)$ in an optimistic scenario. 
Another piece of information that would be attained by the measurement of $\widetilde A_{\rm CP}(K^0\rightarrow\mu^+\mu^-)$ is its sign. 
Within the SM, the CP asymmetry can have either sign (see Sec.~\ref{sec:SMprediction}). This sign ambiguity is linked to the unknown sign of the amplitude $A(K^0_L\to\gamma\gamma)$, which results in the discrete ambiguity in the SM prediction for ${\cal B}(K^0_L\to\mu^+\mu^-)$ (see Sec.~\ref{sec:sign-absorptive}).
This raises the question of how well LHCb could resolve the sign of the CP asymmetry, corresponding to the sign of $A_{\gamma\gamma} \equiv A(K^0_L\to\gamma\gamma)$, assuming SM values for all other parameters (including $\bar\eta$). The results of our study are shown in Fig.~\ref{fig:exp_results} right. 

\begin{figure}[t]
    \includegraphics[width=0.6\textwidth]{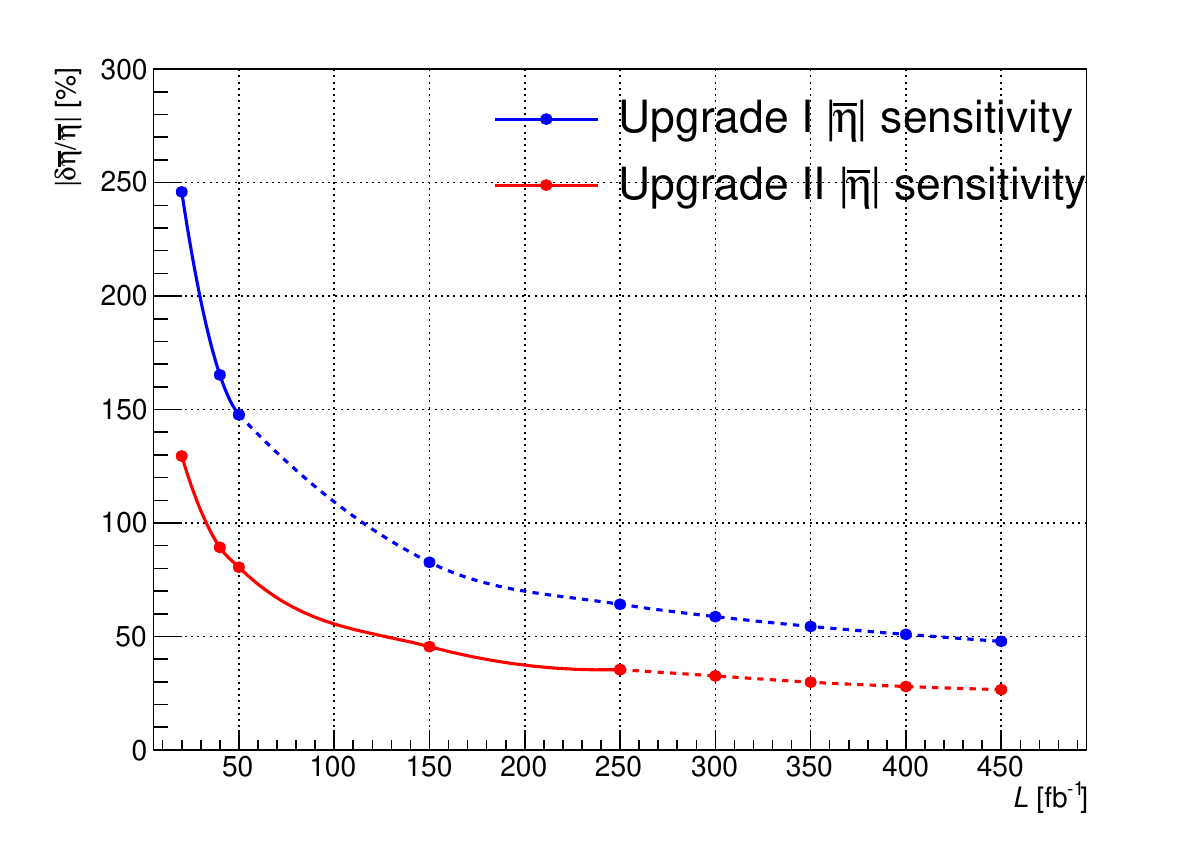}
    \caption{Sensitivity to  $\vert\bar{\eta}\vert$ as a function of the integrated luminosity for Upgrade~I (Upgrade~II) scenarios with $T_P=22\%$, $S_{eff}^{DD}/S_{eff}^{LL} = 0(1.9)$ and a $S^2/(S+B)$ improvement factor of $\times9$ (as in the second and last columns of Tab~\ref{tab:yeffs}).
    Solid lines correspond to currently considered expected luminosities, while the dashed lines demonstrate the behavior outside of the current plans and serve to compare between the performances in the two considered scenarios. 
    \label{fig:exp_results2}}
\end{figure}

Finally, we estimate the sensitivity of the LHCb Upgrades as a function of the integrated luminosity in an optimistic scenario. We assume $T_P=22\%$, an $S^2/(S+B)$ improvement factor of $\times9$, and use $S_{eff}^{DD}/S_{eff}^{LL}=0$ for Upgrade I and $S_{eff}^{DD}/S_{eff}^{LL}=1.9$ for Upgrade II. 
The obtained $|\bar{\eta}|$ sensitivity curves for Upgrade I (blue) and II (red) are shown in Fig.~\ref{fig:exp_results2}.

The presented study demonstrates that a tagged analysis offers an exciting opportunity to measure the $K^0 \rightarrow  \mu^+\mu^+$ decay parameters, with achievable uncertainties for $|\bar{\eta}|$ of about~35\% under the LHCb Upgrade II scenario. The uncertainty of $\widetilde{A}_{CP}$ plays a significant role for that, see our estimates in Appendix~\ref{sec:appendix}. It should be noted that the above analysis is based on fast simulations, representing a simplified picture. 

The takeaway messages from the presented analysis are:
\begin{itemize}
    \item The large statistics that could be collected by Upgrade II offer a unique opportunity for a direct measurement of the CP-violating parameter $|\bar{\eta}|$ with an uncertainty of $\sim 35\%$.
    \item The improvement of the acceptance enabled by the planned Upstream Pixel detector is essential to reach the best sensitivity to $|\bar{\eta}|$.
    \item Significant improvements on background rejection are needed, but seem possible.
    \item An excellent flavor tagging is essential for this measurement. According to simulations this can be achieved thanks to most kaons being produced in softer collisions than $b\bar{b}$ pairs. 
\end{itemize}

\section{Conclusion}
%
The outstanding particle yields of the current era, together with upgraded detector capabilities, open the door to new directions in rare kaon decays.
In this work we study the prospects and implications of a future measurement of the CP asymmetry in $K\to\mu^+\mu^-$, assuming certain benchmark performance figures within LHCb.

We find that the CP asymmetry, $\widetilde A_{\rm CP}(K^0\to\mu^+\mu^-)$, is a powerful probe of the short-distance parameters of the $K^0\to\mu^+\mu^-$ decay. This statement relies on the assumption that knowledge of the $K^0_S$ branching ratio, ${\cal B}(K^0_S\to\mu^+\mu^-)$, is acquired, either by a separate analysis, or within the same study by virtue of time-dependent information.
Regardless of a measurement of ${\cal B}(K^0_S\to\mu^+\mu^-)$, we point out that a determination of the sign of $\widetilde A_{\rm CP}(K^0\to\mu^+\mu^-)$ directly determines the relative sign between the short-distance and long-distance contributions to $K^0_L\to\mu^+\mu^-$, thus eliminating the current discrete ambiguity in the SM prediction for ${\cal B}(K^0_L\to\mu^+\mu^-)$.

Our main results are summarized in Figs.~\ref{fig:exp_results} and~\ref{fig:exp_results2}.
We find that the parameter combination $|A^2\lambda^5\bar\eta|$ could be determined by LHCb at the level of $\sim 35\%$,  assuming the SM, in the most optimistic scenario. 
We further find that, assuming the SM, the sign ambiguity in the amplitude $A(K^0_L\to\gamma\gamma)$ could be resolved with a significance of over $3\sigma$.

The optimistic scenario considered for these estimations assumes that $K^0\rightarrow \mu^+\mu^-$ decay vertices can be reconstructed from downstream tracks only without loosing much vertex resolution (which in principle could be possible with the Upstream Pixel), a tagging power of about $22\%$ (which looks feasible according to Pythia) and significant improvements in the rejection of $K_S^0\rightarrow \pi^+\pi^-$ background events. Less optimistic scenarios would still bring very interesting levels of precision.

The measurement described here would be complementary to the effort to measure the same CKM parameter combination via the decay $K^0_L\to\pi^0\nu\overline\nu$, carried out by the KOTO and proposed KOTO II experiments~\cite{KOTO:2024zbl,KOTO:2025gvq}.
These findings further emphasize the role of LHCb as a key player at the forefront of the current kaon program, and call for dedicated studies of the experimental requirements for the proposed analysis within LHCb, as detailed in Sec.~\ref{sec:LHCb}.

\begin{acknowledgements}
We thank Martin Hoferichter and Bai-Long Hoid for useful discussions.
A.D acknowledges support by the Weizmann Institute of Science Women's postdoctoral career development award. Y.G. is supported in part by the NSF grant PHY--2309456.
The work of T.K. is supported by the JSPS Grant-in-Aid for Scientific Research Grant No.\,24K22872 and 25K07276. 
S.S. is supported by the STFC through an Ernest Rutherford Fellowship under reference ST/Z510233/1 and the grant ST/X003167/1. D.M.S is supported by XuntaGal and by grant PID2022-139514NB-C32 from Minisiterio de Ciencia , innovaci\'on y Universidades (Spain).
\end{acknowledgements}

\appendix

\section{Analytic estimation of sensitivity to the time-integrated CP asymmetry \label{sec:appendix}}
In order to gain an analytic understanding of the sensitivity derived in Fig.~\ref{fig:exp_results} left from a fit to simulated data, we consider a simplified counting experiment analysis.
We assume Poisson errors on the observed number of $K^0$-tagged and $\Kb^0$-tagged decays, $N_{K^0}^{\rm obs.}$ and $N_{\Kb^0}^{\rm obs.}$, and derive the analytic relative error on the integrated CP asymmetry, as a function of the expected yield.

\begin{figure}[t]
    \includegraphics[width=0.6\textwidth]{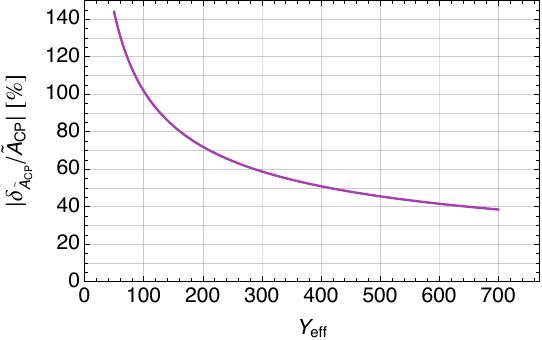}
    \caption{Relative uncertainty on $\widetilde A_{\rm CP}$ as a function of the effective yield, taking into account only statistical errors on the observed yields, as in Eq.~\eqref{eq:ACPrelErrAlt}. We assume the scenario $\sin\varphi_0>0,\, \cos\varphi_0>0$. We take the efficiency function as in Eq.~\eqref{eq:FeffApprox1} with $\beta_{acc.}=4.1\, {\rm ns}^{-1}$ (see Tab.~\ref{tab:acceptances}).
    For the integration range, we use $t_{\min}=0.1\tau_S$, $t_{\text{max}}=2\,\tau_S$ for the calculation of $\widetilde A_{\rm CP}$, and scale $Y_{eff} = Y^\prime_{eff}(t_{\max}=2\,\tau_S)/0.62$ to be consistent with Fig.~\ref{fig:exp_results}.
    We use the predicted SM central value for the parameters $C_S$ and $C_{\rm Int.}$ for this estimate. 
    }
    \label{fig:sigmaACP}
\end{figure}

The relative uncertainty on 
\beq \label{eq:ACPsimp}
\widetilde A_{\rm CP}= \frac{N_{\Kb^0}^{\rm obs.}-N_{K^0}^{\rm obs.}}{N_{\Kb^0}^{\rm obs.}+N_{K^0}^{\rm obs.}}\,,
\eeq
is then given by
\begin{equation} \label{eq:ACPrelErr}
    \left|\frac{\delta \widetilde A_{\rm CP}}{\widetilde A_{\rm CP}}\right| = \sqrt{\frac{4N_{\Kb^0}^{\rm obs.}N_{K^0}^{\rm obs.}}{(N_{\Kb^0}^{\rm obs.}+N_{K^0}^{\rm obs.})(N_{\Kb^0}^{\rm obs.}-N_{K^0}^{\rm obs.})^2}}\, .
\end{equation}
Using Eq.~\eqref{eq:ACPsimp} and
\begin{equation}\label{eq:definition-Yeff}
    Y_{\rm eff.} = N_{K^0}^{\rm obs.}+N_{\Kb^0}^{\rm obs.}\,, 
\end{equation}
we can rewrite Eq.~\eqref{eq:ACPrelErr} as
\begin{equation} \label{eq:ACPrelErrAlt}
    \left|\frac{\delta \widetilde A_{\rm CP}}{\widetilde A_{\rm CP}}\right| = \sqrt{\frac{1-\widetilde A_{\rm CP}^2}{\widetilde A_{\rm CP}^2\,Y_{eff.}}}\, .
\end{equation}

We note that integrating the CP asymmetry beyond $t\sim 2\,\tau_S$ results in a dilution of the asymmetry, since this is the approximate lifetime of the interference effect.
We therefore do not use the full available range for Upgrade II, $t_{\rm max} = 500\, \text{ps} = 5.58\,\tau_S$, but rather cut the integration off at $t_{\rm max}^\prime = 2\, \tau_S$.
In order to be able to compare with the results of Sec.~\ref{sec:results}, we then scale the effective yield accordingly. We find that 
\begin{equation}
    Y_{eff}^\prime(t_{\max}=2\,\tau_S) = 0.62 \cdot Y_{eff}(t_{\max} = 5.58\,\tau_S)\, . 
\end{equation}

The expected sensitivity is plotted in Fig.~\ref{fig:sigmaACP}, as a function of the total effective yield.
We find that the sensitivity to $\widetilde A_{\rm CP}$ is of the same order to that of $|\bar\eta|$, as in the left panel of Fig.~\ref{fig:exp_results}.
This suggests that the statistical error in the measurement of the integrated $\widetilde A_{\rm CP}$ plays a significant part in the sensitivity of the full analysis. The inclusion of shape (time-dependent) information in the full numerical analysis is able to improve on this simple counting experiment estimate.

\bibliographystyle{utphys28mod}
\bibliography{refs}

\end{document}